\documentclass[a4paper]{article}

\usepackage{INTERSPEECH2020}
\usepackage{cite}
\usepackage{bm}
\usepackage{multirow}
\usepackage{xcolor}

\title{Multi-Encoder-Decoder Transformer for Code-Switching Speech Recognition}

\name{Xinyuan Zhou$^{1,2}$, Emre Y{\i}lmaz$^2$, Yanhua Long$^{1}$\thanks{Xinyuan Zhou is an intern at National University of Singapore. Yanhua Long is the corresponding author. The work is supported by the National Natural Science Foundation of China (No.61701306 ), the National Research Foundation Singapore under its AI Singapore Programme (Award Number: AISG-100E-2018-006), and by the National Research Foundation Singapore under the National Robotics Programme, Human-Robot Interaction Phase 1 (Grant No. 192 25 00054).}, Yijie Li$^3$, Haizhou Li$^2$}

\address{
  $^1$Shanghai Normal University, Shanghai, China\\
  $^2$National University of Singapore, Singapore\\
  $^3$Unisound AI Technology Co., Ltd., Beijing, China}
\email{xinyuan\_zhou@u.nus.edu, emrey@kth.se, yanhua@shnu.edu.cn, liyijie@unisound.com, haizhou.li@nus.edu.sg}

\begin{document}

\maketitle
\begin{abstract}
\label{abs}

Code-switching (CS) occurs when a speaker alternates words of two or more languages within a single sentence or across sentences.  Automatic speech recognition (ASR) of CS speech has to deal with two or more languages at the same time. In this study, we propose a Transformer-based architecture with two symmetric language-specific encoders to capture the individual language attributes, that improve the acoustic representation of each language. These representations are combined using a language-specific multi-head attention mechanism in the decoder module. Each encoder and its corresponding attention module in the decoder are pre-trained using a large monolingual corpus aiming to alleviate the impact of limited CS training data. We call such a network a multi-encoder-decoder (MED) architecture. 
Experiments on the SEAME corpus show that the proposed MED architecture achieves $10.2\%$ and $10.8\%$ relative error rate reduction on the CS evaluation sets with Mandarin and English as the matrix language respectively.

\end{abstract}


\section{Introduction}

Code-switching (CS) or code-mixing is defined as the language alternation in an utterance or discourse\cite{bullock2009cambridge,auer2013code}. CS occurs commonly in everyday conversations in multilingual societies. For example, Mandarin and English are often mixed in Singapore and Malaysia, while Cantonese and English are mixed in colloquial Cantonese in Hong Kong~\cite{li2000cantonese}. Most commercial ASR systems, however, are designed to recognize one language, that limits the scope of the applications. To handle CS speech, there have been studies in acoustic modeling~\cite{vu2012,wu2015}, language modeling\cite{li2012, grandee}, and ASR systems~\cite{yilmaz2016_2,westhuizen2017}. 

Recently, the sequence-to-sequence acoustic modeling has attracted great attention in ASR research~\cite{chan2016listen,chiu2018state}. Unlike the conventional hybrid ASR framework,  this architecture encapsulates the acoustic and language information jointly in a single network. The recently proposed Transformer-based end-to-end ASR architectures use deeper encoder-decoder architecture with feedforward layers and multi-head attention for sequence modeling\cite{vaswani2017attention}, and comes with the advantages of parallel computation and capturing long-contexts without recurrence~\cite{dong2018speech,chen2019non,shiliang2019,Transformer-CTC}. These networks have provided comparable performance to the conventional hybrid and other end-to-end approaches on several well-known benchmarks~\cite{karita2019}.

CS ASR is typically a low-resource task due to the scarce acoustic and text resources that contain CS. In the multilingual ASR, to solve the data scarcity problem during acoustic modeling~\cite{ghoshal2013multilingual,huang2013cross}, one could adapt a well-trained high-resource language acoustic model to the target low-resource domain using transfer learning~\cite{ghoshal2013multilingual,huang2013cross, vu2013multilingual,yue}. Similarly in CS ASR, we can adapt two well-trained acoustic models towards a low-resource code-switching scenario. Another strategy is apply multi-task learning technique to exploit the language-specific attributes and alleviate the data sparsity in CS ASR, by jointly training the ASR acoustic model and the language identification classifier~\cite{lyu2008language,zeng2018end,toshniwal2018multilingual}. In these works, the language-specific information was only captured in the decoder or the deep neural network output layer. 

To construct a CS ASR system, there are basically two groups of thoughts. One is to train two language dependent acoustic-language models independently and optimize the combination of them for CS speech\cite{zhang2019towards}. Another is to train a single bilingual acoustic-language model for two languages\cite{yilmaz2016_2,toshniwal2018multilingual}. The former benefits from the precise language specific modeling, but has to face a challenge as to how to integrate two decoders of different probability distributions for a single unbiased decoding process; the latter benefits from the unified training framework that leads to a single bilingual model for two languages, however it is optimized to the average of two very different languages, that cannot be the best for either one.

%Motivated by the promising performance of the Transformer-based ASR systems on monolingual and multilingual ASR scenarios, 
We propose a modified Transformer architecture that takes the best of the two groups of thought. Particularly, we introduce a multi-encoder-decoder (MED) Transformer architecture with two language-specific symmetric encoder branches and corresponding attention modules in the decoder. While MED benefits from the language specific encoder and decoder, it employs a multi-layer decoder structure that unifies the decoding.  Each of the encoder and attention module is pre-trained using a large monolingual corpus for effective initialization.  The complete MED-Transformer model is finally optimized using a smaller amount of code-switching data with Connectionist Temporal Classification (CTC)\cite{graves2006connectionist,watanabe2017hybrid} and Kullback-Leibler divergence\cite{szegedy2016rethinking} criteria. 
%Byte Pair Encoding (BPE)-based~\cite{sennrich2015neural} subwords for English and Chinese characters for Mandarin are used as language-universal modeling units. 
Experiments on the Mandarin-English CS corpus, SEAME, show that the proposed pre-trained MED-Transformer does not only effectively exploit the discrimination between the mixed languages, but also alleviates the CS training data scarcity problem to a certain extent.

%% EY: This can be commented out if we eventually need to open up some space.
The rest of the paper is organized as follows. Section \ref{sec:trans} details the fundamentals of the Transformer architecture. Section \ref{sec:med} describes the MED-Transformer model proposed in this paper. The experimental setup is described in Section \ref{sec:exp}. The recognition results are presented and discussed in Section \ref{sec:rst}.

\section{Transformer Architecture}
\label{sec:trans}

\subsection{Encoder-Decoder with Attention}
\label{subsec:ed}

Like many neural sequence transduction models, the Transformer model also uses an encoder-decoder architecture~\cite{vaswani2017attention,dong2018speech,Transformer-CTC}. In this architecture, the encoder can be regarded as a feature extractor, which converts the input vector $\bm{x}$ into a high-level representation $\bm{h}$.
Given $\bm{h}$, the decoder generates prediction sequence $\bm{y}$
one token at a time auto-regressively. In ASR tasks, tokens are usually modeling units, such as phones, characters or sub-word. During the decoding, the output token at the previous time step, $y_{t-1}$, is also taken as an input to predict the output $y_t$.

The encoder consists of $N$ layers, each of which contains two sub-layers: (1) a multi-head self-attention (MHA) and (2) a position-wise fully connected feed-forward network (FFN). Similar to the encoder, the decoder is also composed of a stack of $M$ identical layers. In addition to the two sub-layers, each layer of decoder also has a third sub-layer between the FFN and MHA, which performs multi-head source-target attention over the output representation of the encoder stack.

\subsection{Multi-Head Attention}
\label{subsec:mhsa}

Multi-head attention is the core module of the Transformer. Unlike ordinary attention, MHA can learn the relationship between queries, keys and values from different subspaces. It takes the ``Scaled Dot-Product Attention" with the following form:
\begin{equation}
  \text{Attention}(Q,K,V) = \text{softmax}(\frac{QK^T}{\sqrt{d_k}})V,
  \label{eq1}
\end{equation}
where $Q \in \mathbb{R}^{t_{q} \times {d_q}}$ are the queries, $K \in \mathbb{R}^{t_{k} \times {d_k}}$ are the keys and $V \in \mathbb{R}^{t_{v} \times {d_v}}$ are the values with input length $t_*$ and the dimension of corresponding elements $d_*$. To prevent pushing the softmax into extremely small gradient regions caused by large dimensions,
the $\frac{1}{\sqrt{d_k}}$ is used to scale the dot products.

In order to calculate attention from multiple subspaces, multi-head attention is constructed as follows:
\begin{equation}
\text{MHA}(Q,K,V) = \text{Concat}(Head_1, ..., Head_H)W^O, \label{eq2}
\end{equation}
\begin{equation}
\qquad Head_i = \text{Attention}(QW_i^Q, KW_i^K, VW_i^V), \label{eq3}
\end{equation}
where $W_i^*$ is the projection matrices with $W_i^Q \in \mathbb{R}^{d_{\text{model}} \times {d_Q}}$, $W_i^K \in \mathbb{R}^{d_{\text{model}} \times {d_K}}$, $W_i^V \in \mathbb{R}^{d_{\text{model}} \times {d_V}}$ and $W^O \in \mathbb{R}^{d_{\text{model}} \times {d_O}}$, $d_{\text{model}}$ is the dimension of the input vector to the encoder.

$Q$, $K$ and $V$ in each attention are projected to $d_*$-dimensional space through three linear projections layers $W_i^*$ respectively. After performing $H$ attentions, namely calculating the representations at $H$ different subspaces, the output values are concatenated and projected using  $W^O$ to get the MHA output.

\subsection{Positional Encoding}
\label{subsec:pe}

The multi-head attention contains no recurrences and convolutions unlike the recurrent and convolutional neural networks, hence, it cannot model the order of the input acoustic sequence. For this reason, ``positional encodings" are used to learn the positional information in the input speech. The formulae of the positional encoding are shown as follows:
\begin{align}
\text{PE}_{(\text{pos},2i)} &= \text{sin}(\text{pos}/10000^{2i/d_{\text{model}}}), \label{eq4} \\
\text{PE}_{(\text{pos},2i+1)} &= \text{cos}(\text{pos}/10000^{2i/d_{\text{model}}}), \label{eq5}
\end{align}
where $\text{pos}$ is the position of the current frame/token in the current utterance/text and $i$ is the dimension. The encoding values at the even and odd positions are calculated
by equation (\ref{eq4}) and (\ref{eq5}) respectively. By adding different values to different dimensions of each input element, the position information is integrated into the attention process. In this way, the model can learn the relative position between different elements.

\section{Multi-Encoder-Decoder Transformer}
\label{sec:med}

\begin{figure}[t]
  \centering
  \includegraphics[width=\linewidth]{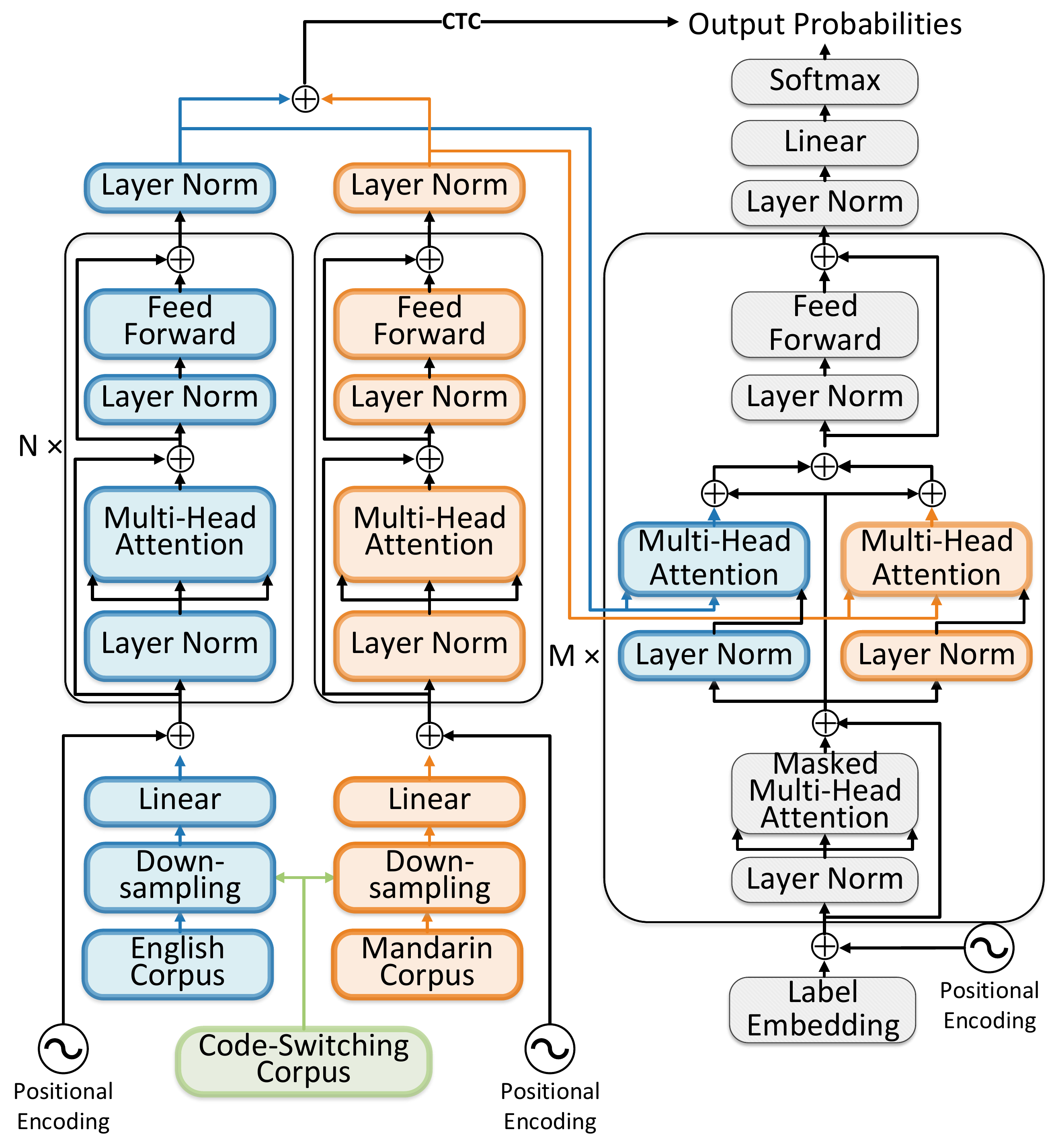}
  \caption{Model architecture of the proposed multi-encoder-decoder Transformer. }
  \label{fig:med}
\end{figure}

\subsection{Architecture}
\label{subsec:art}
The detailed model architecture of the proposed MED Transformer is illustrated in Figure \ref{fig:med}. Similar to the standard Transformer, the proposed model is composed of a stack of encoder and decoder layers. The main difference is that two language-specific encoder modules, which are marked in blue for English and orange for Mandarin, are incorporated to learn the individual language attributes. The goal is to enhance the discrimination between the high-level representations given at the output of each encoder by using separate components for each mixed language. 
And to well integrate these high-level discriminative representations together with the acoustic and context code-switching property for CS ASR. The outputs of these two encoders are combined using a language-specific multi-head self-attention mechanism in the decoder layer. 

The structure of the language-specific MHA is shown in the right panel of Figure \ref{fig:med}. This language-specific MHA procedure can be formulated as follows:
\begin{align}
\text{MidLyr} &= \frac{RC^{\text{Eng}}+RC^{\text{Man}}}{2},        \label{eq6} \\
RC^* &= \text{UndLyr}+\text{MHA}^*(Q^*, K^*, V^*), \label{eq7} \\
Q^* &= \text{LayerNorm}^*(\text{UndLyr}),\\
K^* &= V^* = \text{Encoder}^*(x),     \label{eq8}
\end{align}
where $* \in \{\text{Eng}, \text{Man}\}$, $RC^* \in \mathbb{R}^{l \times {d_{\text{model}}}}$ is the corresponding residual connection\cite{he2016deep} which is added by the output of under layer ($\text{UndLyr}$) and $\text{MHA}^*$, $l$ is the length of input token sequence. The mean $RC$ values, MidLyr, are fed into the next sub-layer.

Intra-language transfer learning, i.e., each encoder is first trained on vast amounts of monolingual resources and then fine-tuned using the CS resources, is applied to remedy the lack of a large CS corpus. In the proposed architecture, the language specific encoders and the corresponding attention modules are pre-trained using a large monolingual English and Mandarin corpus respectively. Using the pre-trained parameters for initialization, we fine-tune the MED Transformer model on the code-switching speech corpus by jointly training with respect to the CTC and KLD criteria.

The downsampling block in our model is the same as described in \cite{Transformer-CTC}. It is performed by using two $3\times3$ CNN layers with stride 2 to reduce the GPU memory occupation and the length difference from the label sequence.

\subsection{Multi-Objective Learning}
\label{subsec:ctc}

As described in \cite{watanabe2017hybrid,Transformer-CTC}, in order to benefit from the monotonic alignment, the CTC loss function is used to jointly train the encoder of the proposed MED Transformer in a multi-objective learning style:
\begin{equation}
\mathcal{L}_{\text{MOL}} = \lambda \mathcal{L}_{\text{CTC}} + (1-\lambda)\mathcal{L}_{\text{Attention}} \label{eq10}
\end{equation}
with a tuning parameter $\lambda\in[0,1]$. The two language-specific encoders are trained simultaneously by calculating the CTC loss function of the sum of both encoders' output. We call the multi-objective learning with CTC loss function as MOL in the experiments.

\section{Experiments}
\label{sec:exp}

\subsection{Corpora}
\label{subsec:cp}

Three corpus are used in our experiments:~(1) AISHELL-2~\cite{aishell2} with 1K hours monolingual Mandarin read speech, (2) Librispeech~\cite{panayotov2015librispeech} with 960 hours monolingual English read speech, and (3) SEAME\cite{lyu2010} with 112 hours Mandarin-English code-switching conversational speech.~These two monolingual datasets of almost equal size are used for the language-specific model components pre-training. In the code-switching corpus, the SEAME contains not only CS utterances, but also a small amount of monolingual utterances as summarized in Table \ref{tab:SEAME1}. 

\begin{table}[!htbp]
   \caption{The composition of Mandarin, English monolingual speech, and Mandarin-English CS speech in SEAME corpus.}
  \label{tab:SEAME1}
  \centering
  \begin{tabular}{c c c c }
    \toprule
    & \textbf{Mandarin} & \textbf{English} & \textbf{Code-Switching}\\
    \midrule
    Ratio       & 19.6\%      & 18.9\%       & 61.5\%  \\
    \#utterances  & 31,809 & 30,673 & 99,808  \\
    \bottomrule
  \end{tabular}
\end{table}

To fairly evaluate the CS performance of ASR systems, two evaluation sets were designed in the SEAME corpus: (1) eval\textsubscript{man} contains CS speech with Mandarin as the matrix language and (2) eval\textsubscript{sge} contains CS speech with Singapore English as the matrix language. For development purposes, we randomly selected 5\% of the training utterances (3.4 hours) as the development set. The details of the training, development and evaluation sets are shown in Table \ref{tab:SEAME2}.

\begin{table}[t]
  \caption{The division of training, development, and evaluation data of SEAME corpus.}
  \label{tab:SEAME2}
  \centering
  \begin{tabular}{c c c c c}
    \toprule
    &train          &dev      & eval\textsubscript{man}  &eval\textsubscript{sge}\\
    \midrule
    Speakers        &132        &5          &10             &10     \\
    Duration (hrs)  &97.6       &3.4        &7.5           &3.9   \\
    Mandarin (\%)   &59         &52         &69             &29   \\
    \bottomrule
  \end{tabular}
\end{table}

\subsection{Experimental Setup}
\label{subsec:exp}
All of the experiments are implemented and performed using the ESPnet~\cite{watanabe2018espnet} end-to-end speech processing toolkit.
We extract 80-dimensional log Mel-filterbank plus pitch and its $\Delta$, $\Delta\Delta$ as acoustic features and normalize them with global mean computed using the training set. The frame-length is 25 ms with a 10 ms shift. SpecAugment~\cite{park2019specaugment} is applied for data augmentation during all model pre-training and fine-tuning stages. Furthermore, the speed perturbation \cite{ko2015audio} is also used to enhance the model robustness during the fine-tuning.

The MED Transformer in our experiment contains 12-layer encoder and 6-layer decoder, where the $d_{\text{model}}=256$ and the dimensionality of inner-layer in FFN $d_{\text{ff}}=2048$. In all attention sub-layers, 4 heads are used for MHA. 
And as for CTC, hyperparameters $\lambda=0.3$. The whole network is trained for 30 epochs and warmup\cite{vaswani2017attention} is used for first 25K iterations.

We use sub-words instead of letters as the Transformer modeling units for English, and Chinese characters for Mandarin. 5,000 subwords were generated from Librispeech using BPE. 4,230 Chinese characters are extracted from the AISHELL-1 database. These subwords and characters are then combined together to form the 9,230 Character-BPE modeling units.

Ten-hypotheses-width beam search is used with the the one-pass decoding for CTC as described in \cite{watanabe2017hybrid} and a two-layer RNN language model (LM) shallow fusion\cite{shallowfusion}, which was trained on the training transcriptions of SEAME with 2,574 Chinese characters and 12,822 English words. We use the token error rate (TER) as the ASR performance measure. The ‘token’ here refers to the unit of Mandarin character and English word, respectively. 
\begin{table}[t]
  \caption{A comparative study of system performance (TER\%) among various Speech-Transformer configurations, with multi-objective learning (MOL) or LM shallow fusion.}
  \label{tab:BASELINE}
  \centering
  \begin{tabular}{l l l l l}
    \toprule
    &Model                &eval\textsubscript{man}   &eval\textsubscript{sge}  \\
    \midrule
    &Kaldi (LF-MMI)                    & 19.0          & 26.6                \\
    &Transformer                       & 25.2          & 33.7                \\
    &Transformer+MOL                   & 18.9          & 26.2                \\
    &Transformer+MOL+LM                & 18.6          & 25.9                \\
 %   &Transformer+MLT+LM+pretrain       & 19.3          & 27.1                \\

    \bottomrule
  \end{tabular}
\end{table}
\subsection{Comparative Study}
\label{subsec:base}

We conduct a number of experiments in an ablation study. Transformer has been successfully \cite{dong2018speech}  applied the in ASR, that is called ``Speech-Transformer". First, we would like to see how the baseline Transformer with a single encoder-decoder pipeline performs, we then study the effect of  adding CTC loss function in the multi-objective learning~\cite{kim2017joint,Transformer-CTC}, and how the LM shallow fusion improves the performance as a post-processing. 
%The 9,230 Character-BPE units were taken as the Transformer modeling units. The performance of the standard Speech-Transformer is improved by applying \textcolor{red}{CTC and LM rescoring}. 

For ease of comparison, we include the previously reported results~\cite{guo2018study}. It was obtained with the state-of-the-art LF-MMI-based hybrid Kaldi system \cite{lfmmi} with a trigram LM. The results given by all baseline systems with single encoder-decoder pipeline are presented in Table \ref{tab:BASELINE}. 

From these results, we can conclude the Transformer with CTC joint training outperforms the standard Speech-Transformer and the LF-MMI based hybrid system. The ASR performance of Transformer+MOL can be further improved by performing LM shallow fusion. Applying pre-training to the Transformer+MOL+LM system, in which the standard Transformer architecture is initialized on the combination of two large monolingual corpora, does not bring any improvements. Therefore, we use the Transformer+MOL+LM system as our baseline.

The proposed MED Transformer contains two language-specific encoders and MHA layers in the decoder. To investigate the impact of each individual language-specific module on the ASR performance, we report the performance of two additional systems: (1) Transformer (M-En) with a standard decoder and two language-specific encoders and (2) Transformer (M-De) with a standard encoder and two language-specific MHA layers in the decoder.

\section{Results and Discussion}
\label{sec:rst}

Table \ref{tab:result} shows the detailed TER results on SEAME given by the proposed MED Transformer-based ASR system. The proposed MED Transformer provides a relative 10.2\% TER reduction on the eval\textsubscript{man} set, and 10.8\% relative reduction on the eval\textsubscript{sge} set compared to the baseline system. When we compare the results on English and Mandarin part of both evaluation sets, it can be seen that the recognition of the English words are more challenging than the Mandarin characters. MED Transformer yields improvements on both English words and Mandarin characters. Specifically, the relative TER reductions are 12.8\% and 12.1\% on English words and 8.6\% and 6.2\% on Mandarin characters in the eval\textsubscript{man} set and eval\textsubscript{sge} set, respectively. 

\begin{table}[t]
\caption{TER\% results on SEAME with different Transformer-based ASR architectures.
`All', `Man', `Eng' represent the TER results on the complete evaluation set, Mandarin characters and English words, respectively. All systems are with MOL+LM.}
\label{tab:result}
\centering
\begin{tabular}{lllllll}
\toprule
\multirow{2}{*}{System} & \multicolumn{3}{c}{eval\textsubscript{man}}  & \multicolumn{3}{c}{eval\textsubscript{sge}} \\
\cline{2-4} \cline{5-7}
            &\textbf{All}    & Man    &Eng   &\textbf{All}   &Man    &Eng     \\
  \midrule
  Baseline &    18.6  &   15.1 & 29.0   & 25.9    &  19.5  & 29.7  \\

  M-En     &    16.8  &   13.8 & 25.5   & 23.4    &  18.4  & 26.3 \\
  M-De     &    19.1  &   15.5 & 29.7   & 26.5    &  20.0  & 30.3 \\
  \midrule
  MED      &    16.7  &   13.8 & 25.3   & 23.1    &  18.3  & 26.1 \\
  \bottomrule
  \end{tabular}
  \vspace{-0.2cm}
\end{table}

In the middle panel of Table \ref{tab:result}, it can be seen that the language-specific encoders do provide better high-level representations to encode the discriminative information for each language. The MED Transformer achieves 16.7\% TER on the eval\textsubscript{man} set,  and 23.1\% TER on eval\textsubscript{sge} set, as we know, these are the best result on SEAME in the literature. The performance of M-En is similar to the performance of the MED system. On the contrary, since the decoders mainly rely on the output of the encoder during model training, using only the language-specific MHAs in the decoder does not improve the CS ASR performance.

\begin{figure}[t]
  \centering
  \includegraphics[width=\linewidth]{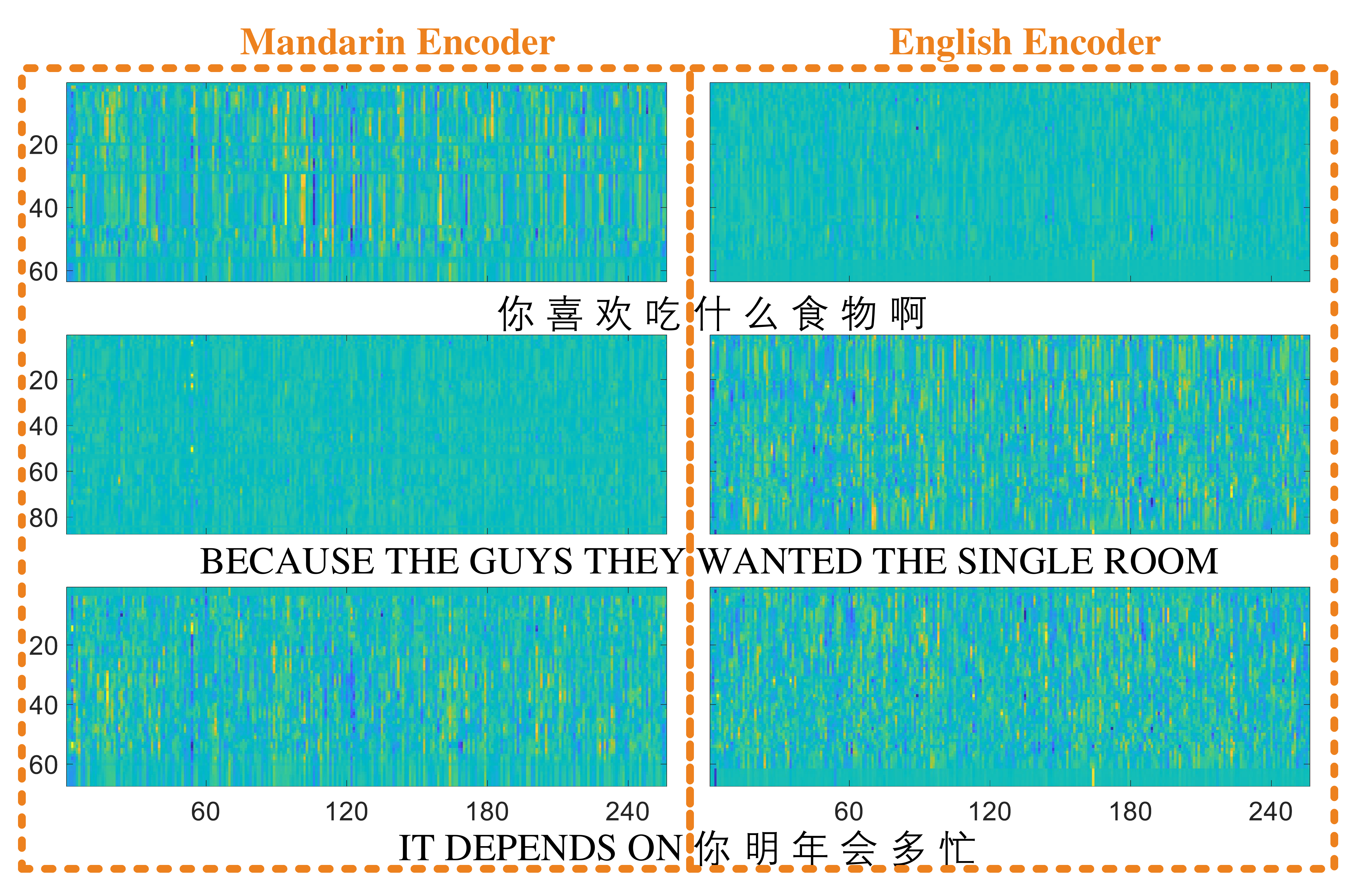}
  \caption{The distribution of the normalized values observed at each encoder's final layer of the proposed MED Transformer system}
  \label{fig:outputdis}
\end{figure}

We examine how the two encoders respond to input speech of different languages, namely monolingual English, monolingual Mandarin and a CS utterance. The distribution of the normalized values observed at each encoder's final layer are given in Figure \ref{fig:outputdis}. The values on the horizontal axis are the attention vector indices and the vertical axis has the frames indices after down-sampling. 

The upper panel shows that the Mandarin encoder has a considerably larger variance at the output than the English encoder, which can be interpreted as Mandarin encoder is responding to Mandarin utterance while English encoder is not. Likewise, the middle panel illustrates a similar pattern for the monolingual English utterance. Both plots in the lower panel shows comparable activation for the CS utterance. These plots gives insight about how the encoders of the MED transformer system behave for different inputs. The differences between the encoder outputs demonstrate how the proposed architecture effectively captures the language-specific information and better discriminate among the mixed languages during the recognition accounting for the improvements reported in Table \ref{tab:result}.

\section{Conclusion}
\label{sec:cls}
This paper investigates a new Transformer-based ASR architecture for the Mandarin-English Code-switching ASR task. Different from the standard Speech-Transformer for monolingual ASR, we proposed a multi-encoder-decoder Transformer to learn the individual language attributes and effectively discriminate among the languages mixed in the target speech. The proposed architecture consists two symmetric language-specific encoders and corresponding language-specific multi-head attention blocks in the decoder module. To alleviate the impact of training data scarcity, we pre-trained all of the language-specific modules in the Transformer using large monolingual speech corpora. 

Experimental results on the SEAME dataset has shown that the proposed architecture outperforms the baseline Transformer system with a single encoder and a state-of-the-art Kaldi-based hybrid ASR system. The investigation of the proposed MED Transformer on a recently released large-scale CS corpus remains as a future work.

\bibliographystyle{IEEEtran}

\bibliography{mybib}

\begin{thebibliography}{10}
\providecommand{\url}[1]{#1}
\csname url@samestyle\endcsname
\providecommand{\newblock}{\relax}
\providecommand{\bibinfo}[2]{#2}
\providecommand{\BIBentrySTDinterwordspacing}{\spaceskip=0pt\relax}
\providecommand{\BIBentryALTinterwordstretchfactor}{4}
\providecommand{\BIBentryALTinterwordspacing}{\spaceskip=\fontdimen2\font plus
\BIBentryALTinterwordstretchfactor\fontdimen3\font minus
  \fontdimen4\font\relax}
\providecommand{\BIBforeignlanguage}[2]{{%
\expandafter\ifx\csname l@#1\endcsname\relax
\typeout{** WARNING: IEEEtran.bst: No hyphenation pattern has been}%
\typeout{** loaded for the language `#1'. Using the pattern for}%
\typeout{** the default language instead.}%
\else
\language=\csname l@#1\endcsname
\fi
#2}}
\providecommand{\BIBdecl}{\relax}
\BIBdecl

\bibitem{bullock2009cambridge}
B.~E. Bullock and A.~J.~E. Toribio, \emph{The Cambridge handbook of linguistic
  code-switching.}\hskip 1em plus 0.5em minus 0.4em\relax Cambridge University
  Press, 2009.

\bibitem{auer2013code}
P.~Auer, \emph{Code-switching in conversation: Language, interaction and
  identity}.\hskip 1em plus 0.5em minus 0.4em\relax Routledge, 2013.

\bibitem{li2000cantonese}
D.~C. Li, ``Cantonese-english code-switching research in hong kong: A y2k
  review,'' \emph{World Englishes}, vol.~19, no.~3, pp. 305--322, 2000.

\bibitem{vu2012}
N.~T. Vu, D.-C. Lyu, J.~Weiner, D.~Telaar, T.~Schlippe, F.~Blaicher, E.-S.
  Chng, T.~Schultz, and H.~Li, ``A first speech recognition system for
  {Mandarin-English} code-switch conversational speech,'' in \emph{Proc.
  ICASSP}, March 2012, pp. 4889--4892.

\bibitem{wu2015}
C.~H. Wu, H.~P. Shen, and C.~S. Hsu, ``Code-switching event detection by using
  a latent language space model and the delta-{Bayesian} information
  criterion,'' \emph{IEEE/ACM Trans. on Audio, Speech, and Lang. Processing},
  vol. 23 (11), pp. 1892--1903, 2015.

\bibitem{li2012}
Y.~Li and P.~Fung, ``Code switching language model with translation constraint
  for mixed language speech recognition,'' in \emph{Proc. COLING}, 2012, pp.
  1671--1680.

\bibitem{grandee}
G.~Lee and H.~Li, ``Modeling code-switch languages using bilingual parallel
  corpus,'' \emph{The 58th annual meeting of the Association for Computational
  Linguistics (ACL)}, 2020.

\bibitem{yilmaz2016_2}
E.~Y{\i}lmaz, H.~Van~den Heuvel, and D.~A. Van~Leeuwen, ``Investigating
  bilingual deep neural networks for automatic speech recognition of
  code-switching {Frisian} speech,'' \emph{Procedia Computer Science}, pp.
  159--166, May 2016.

\bibitem{westhuizen2017}
E.~Van~der Westhuizen and T.~Niesler, ``Synthesising {isiZulu-English}
  code-switch bigrams using word embeddings,'' in \emph{Proc. INTERSPEECH},
  2017, pp. 72--76.

\bibitem{chan2016listen}
W.~Chan, N.~Jaitly, Q.~Le, and O.~Vinyals, ``Listen, attend and spell: A neural
  network for large vocabulary conversational speech recognition,'' in
  \emph{Proc. ICASSP}, 2016, pp. 4960--4964.

\bibitem{chiu2018state}
C.-C. Chiu, T.~N. Sainath, Y.~Wu, R.~Prabhavalkar, P.~Nguyen, Z.~Chen,
  A.~Kannan, R.~J. Weiss, K.~Rao, E.~Gonina \emph{et~al.}, ``State-of-the-art
  speech recognition with sequence-to-sequence models,'' in \emph{Proc.
  ICASSP}, 2018, pp. 4774--4778.

\bibitem{vaswani2017attention}
A.~Vaswani, N.~Shazeer, N.~Parmar, J.~Uszkoreit, L.~Jones, A.~N. Gomez,
  {\L}.~Kaiser, and I.~Polosukhin, ``Attention is all you need,'' in
  \emph{Advances in neural information processing systems}, 2017, pp.
  5998--6008.

\bibitem{dong2018speech}
L.~Dong, S.~Xu, and B.~Xu, ``Speech-transformer: a no-recurrence
  sequence-to-sequence model for speech recognition,'' in \emph{Proc. ICASSP},
  2018, pp. 5884--5888.

\bibitem{chen2019non}
N.~Chen, S.~Watanabe, J.~Villalba, and N.~Dehak, ``Non-autoregressive
  transformer automatic speech recognition,'' \emph{arXiv preprint
  arXiv:1911.04908}, 2019.

\bibitem{shiliang2019}
Z.~Y. Shiliang~Zhang, Ming~Lei, ``Investigation of transformer based spelling
  correction model for ctc-based end-to-end mandarin speech recognition,'' in
  \emph{Proc. INTERSPEECH}, 09 2019, pp. 2180--2184.

\bibitem{Transformer-CTC}
K.~Shigeki, N.~E.~Y. Soplin, S.~Watanabe, D.~Delcroix, A.~Ogawa, and
  T.~Nakatani, ``Improving transformer-based end-to-end speech recognition with
  connectionist temporal classification and language model integration,'' in
  \emph{Proc. INTERSPEECH}, 09 2019, pp. 1408--1412.

\bibitem{karita2019}
S.~{Karita}, N.~{Chen}, T.~{Hayashi}, T.~{Hori}, H.~{Inaguma}, Z.~{Jiang},
  M.~{Someki}, N.~E.~Y. {Soplin}, R.~{Yamamoto}, X.~{Wang}, S.~{Watanabe},
  T.~{Yoshimura}, and W.~{Zhang}, ``A comparative study on transformer vs rnn
  in speech applications,'' in \emph{Proc. IEEE ASRU Workshop}, 2019, pp.
  449--456.

\bibitem{ghoshal2013multilingual}
A.~Ghoshal, P.~Swietojanski, and S.~Renals, ``Multilingual training of deep
  neural networks,'' in \emph{Proc. ICASSP}, 2013, pp. 7319--7323.

\bibitem{huang2013cross}
J.-T. Huang, J.~Li, D.~Yu, L.~Deng, and Y.~Gong, ``Cross-language knowledge
  transfer using multilingual deep neural network with shared hidden layers,''
  in \emph{Proc. ICASSP}, 2013, pp. 7304--7308.

\bibitem{vu2013multilingual}
N.~T. Vu and T.~Schultz, ``Multilingual multilayer perceptron for rapid
  language adaptation between and across language families.'' in \emph{Proc.
  INTERSPEECH}, 2013, pp. 515--519.

\bibitem{yue}
X.~Yue, G.~Lee, E.~Yılmaz, F.~Deng, and H.~Li, ``End-to-end code-switching asr
  for low-resourced language pairs,'' in \emph{Proc. IEEE ASRU Workshop}, 2019,
  pp. 972--979.

\bibitem{lyu2008language}
D.-C. Lyu and R.-Y. Lyu, ``Language identification on code-switching utterances
  using multiple cues,'' in \emph{Proc. INTERSPEECH}, 2008.

\bibitem{zeng2018end}
Z.~Zeng, Y.~Khassanov, V.~T. Pham, H.~Xu, E.~S. Chng, and H.~Li, ``{On the
  end-to-end solution to Mandarin-English code-switching speech recognition},''
  in \emph{Proc. INTERSPEECH}, 2019, pp. 2165--2169.

\bibitem{toshniwal2018multilingual}
S.~Toshniwal, T.~N. Sainath, R.~J. Weiss, B.~Li, P.~Moreno, E.~Weinstein, and
  K.~Rao, ``Multilingual speech recognition with a single end-to-end model,''
  in \emph{Proc. ICASSP}, 2018, pp. 4904--4908.

\bibitem{zhang2019towards}
S.~Zhang, Y.~Liu, M.~Lei, B.~Ma, and L.~Xie, ``Towards language-universal
  mandarin-english speech recognition,'' \emph{Proc. INTERSPEECH}, pp.
  2170--2174, 2019.

\bibitem{graves2006connectionist}
A.~Graves, S.~Fern{\'a}ndez, F.~Gomez, and J.~Schmidhuber, ``Connectionist
  temporal classification: labelling unsegmented sequence data with recurrent
  neural networks,'' in \emph{Proceedings of the 23rd international conference
  on Machine learning}, 2006, pp. 369--376.

\bibitem{watanabe2017hybrid}
S.~Watanabe, T.~Hori, S.~Kim, J.~R. Hershey, and T.~Hayashi, ``Hybrid
  ctc/attention architecture for end-to-end speech recognition,'' \emph{IEEE
  Journal of Selected Topics in Signal Processing}, vol.~11, no.~8, pp.
  1240--1253, 2017.

\bibitem{szegedy2016rethinking}
C.~Szegedy, V.~Vanhoucke, S.~Ioffe, J.~Shlens, and Z.~Wojna, ``Rethinking the
  inception architecture for computer vision,'' in \emph{Proceedings of the
  IEEE conference on computer vision and pattern recognition}, 2016, pp.
  2818--2826.

\bibitem{he2016deep}
K.~He, X.~Zhang, S.~Ren, and J.~Sun, ``Deep residual learning for image
  recognition,'' in \emph{Proceedings of the IEEE conference on computer vision
  and pattern recognition}, 2016, pp. 770--778.

\bibitem{aishell2}
J.~{Du}, X.~{Na}, X.~{Liu}, and H.~{Bu}, ``{AISHELL-2: Transforming Mandarin
  ASR Research Into Industrial Scale},'' \emph{ArXiv}, Aug. 2018.

\bibitem{panayotov2015librispeech}
V.~Panayotov, G.~Chen, D.~Povey, and S.~Khudanpur, ``Librispeech: an asr corpus
  based on public domain audio books,'' in \emph{Proc. ICASSP}, 2015, pp.
  5206--5210.

\bibitem{lyu2010}
D.-C. Lyu, T.-P. Tan, E.~S. Chng, and H.~Li, ``Seame: a mandarin-english
  code-switching speech corpus in south-east asia,'' in \emph{Proc.
  INTERSPEECH}, Sept. 2010, pp. 1986--1989.

\bibitem{watanabe2018espnet}
S.~Watanabe, T.~Hori, S.~Karita, T.~Hayashi, J.~Nishitoba, Y.~Unno, N.~{Enrique
  Yalta Soplin}, J.~Heymann, M.~Wiesner, N.~Chen, A.~Renduchintala, and
  T.~Ochiai, ``Espnet: End-to-end speech processing toolkit,'' in \emph{Proc.
  INTERSPEECH}, 2018, pp. 2207--2211.

\bibitem{park2019specaugment}
D.~S. Park, W.~Chan, Y.~Zhang, C.-C. Chiu, B.~Zoph, E.~D. Cubuk, and Q.~V. Le,
  ``Specaugment: A simple augmentation method for automatic speech
  recognition,'' in \emph{Proc. INTERSPEECH}, 2019, p. 2613–2617.

\bibitem{ko2015audio}
T.~Ko, V.~Peddinti, D.~Povey, and S.~Khudanpur, ``Audio augmentation for speech
  recognition,'' in \emph{Proc. INTERSPEECH}, 2015, pp. 3586--3589.

\bibitem{shallowfusion}
A.~Kannan, Y.~Wu, P.~Nguyen, T.~N. Sainath, Z.~Chen, and R.~Prabhavalkar, ``An
  analysis of incorporating an external language model into a
  sequence-to-sequence model,'' in \emph{Proc. ICASSP}, 2018, pp. 5824--5828.

\bibitem{kim2017joint}
S.~Kim, T.~Hori, and S.~Watanabe, ``Joint ctc-attention based end-to-end speech
  recognition using multi-task learning,'' in \emph{Proc. ICASSP}, 2017, pp.
  4835--4839.

\bibitem{guo2018study}
P.~Guo, H.~Xu, L.~Xie, and C.~E. Siong, ``Study of semi-supervised approaches
  to improving english-mandarin code-switching speech recognition,'' in
  \emph{Proc. INTERSPEECH}, 2018, pp. 1928--1932.

\bibitem{lfmmi}
D.~Povey, V.~Peddinti, D.~Galvez, P.~Ghahremani, V.~Manohar, X.~Na, Y.~Wang,
  and S.~Khudanpur, ``Purely sequence-trained neural networks for asr based on
  lattice-free mmi.'' in \emph{Proc. INTERSPEECH}, 2016, pp. 2751--2755.

\end{thebibliography}

\end{document}